# A possibility of kinetic energy economy at the transition of a two-dimensional conductor to the superconducting state

I. N. Zhilyaev

*Institute of Microelectronics Technology and High Purity Materials, Russian Academy of Sciences, Chernogolovka, Moscow Region, Russia*

zhilyaev@ipmt-hpm.ac.ru

For a two-dimensional conductor under the conditions of the Bardeen-Cooper-Schrieffer (BCS) mechanism of superconductivity, a factor resulting in a economy in kinetic energy upon the transition to the superconducting state and connected to the zero oscillations of macroscopic quantum oscillator is offered. The economy, in as turn, can result in a greater rigidity of the superconducting state.
PACS: 74.20.-z, 74.72.-h

According to the BCS mechanism, electron pairs with opposite directed impulses and participation of virtual phonons form a superconducting state by continuously scattering between individual electronic states in a certain interval of energy determined by an average energy of phonons near the Fermi-level. For these processes to occur, the interval should involve free states. As a result, the distribution of electron pair states by an impulse even at temperature T = 0 does not break abruptly at the Fermi-level, but is smeared. The smearing of energy according to BCS theory corresponds to the size of a superconducting gap $\Delta$. An electronic system at transition to a superconducting state passes on a condition with a greater kinetic energy, which corresponds to a loss of energy from the view point of an energy state (certainly, this loss is compensated for by the energy of attraction via virtual phonons). Our idea is to economy this loss, by impulse smearing with the help of the ground state of a spatial quantum oscillator which can be present in these structures. Zero oscillations of the oscillator kinetic energy should result in dynamic smearing of electronic states in the impulse space. The population of electronic states is then transformed so, that there appears a range of impulses where both the occupied and free states are present. As estimations show, energy smearing can be of order of maximal observed Δ. At the same time, these oscillations do not influence the electron-phonon interaction and do not destroy the Cooper pairing. Because structures having features of high-temperature superconductors (HTSC) are of greatest interest, estimations will be made using parameters characteristic of these materials.

Let's consider a two-dimensional conductor in the form of a square with the sides d*d = 50*50 Å$^2$. The size d is chosen from those reasons that it was a little bit more the coherence length ξ ≈ 30 Å in HTSC planes as at the smaller sizes Cooper pairing weakens. Along such a conductor, plasma waves having the law of dispersion $1/\omega^2 = 1/(c/\lambda)^2 + 2/(\omega_p)^2$ [ 1 ], where c is the speed of light, $\omega_p = (4\pi n e^2/m)^{1/2}$ is the plasma frequency of a three-dimensional conductor (e and m are charge and mass of an electron) can spread. For the charge density n = 5*10$^{21}$ cm$^3$ typical in HTSC, $\omega_p$ will be 4*10$^{15}$ rad/s. In the plane of a conductor, like in a resonator, independent stationary waves can be exited with wave vectors directed along the sides of the square. As λ, $\lambda_r$ = 2d/r are possible where r is an integer. Because of small sizes of the conductor and, hence, small lengths of waves λ, oscillations can arise only at a frequency close to the limiting $\omega_s = \omega_p/\sqrt{2}$ = 2.8*10$^{15}$rad/s. By analogy with a macroscopic quantum oscillator in the Josephson junction [2], oscillations of such a stationary wave can be regarded as a macroscopic quantum oscillator with a discrete energy spectrum $E_l = \hbar\omega_s(1/2+l)$, where l is an integer, provided the dissipation is small enough.

Let's estimate amplitude of zero oscillations. The ground state of the oscillator is owe to the uncertainty ratio according to which the total energy $E_0 = \hbar\omega_s(1/2)$ is given precisely and components, the kinetic energy of state $E_k$ and potential $E_p$ , each have nonzero dispersions $(\Delta E_k)^2 = (\Delta E_p)^2 = 2 (\hbar\omega_s/4)^2$ [3]. So, the amplitude of zero oscillations of the kinetic energy, close by the value to $\Delta E_k$, is comparable to the total oscillator energy $E_0 = \hbar\omega_s/2$. The energy of zero oscillations of kinetic energy is the sum of energies of charge carriers oscillations of a conductor. Let N be the number of charge carriers in a conductor. Then the kinetic energy of individual electronic states obliged to zero oscillations $\delta\varepsilon$ is equal to $\Delta E_k/N \approx E_0/N$. Let N be the number of electrons in the volume = $50*50*12$ Å$^3$ (assuming the conducting layer thickness equal to 12 Å, a typical lattice parameter value across layers in HTSC). Then, N will be equal to 150 and $\delta\varepsilon \approx E_0/N = \hbar\omega_s/2N = 9*10^{-3}$ eV (this value corresponds to 100K).

Zero oscillations of macroscopic quantum oscillator can take place only at small losses in a conductor. Consider the dissipation channels in our conductor. Charge carries scattering in the bulk and on the surface and electromagnetic interactions with the environment can influence the oscillator state. Consider influence of these factors taking the oscillator parameters into account. It is usually considered that the free path even in a "dirty" limit is about 100 Å. As the size of our structure is smaller, the influence of volume scattering by impurities and defects can be taken negligible. The temperature effect on the oscillatory state of charge carriers in the ground state of a quantum oscillator would be appreciable only at temperatures near 100 K, because the energy of oscillations, per one electron close to $10^{-2}$ eV. Diffusive charge carriers scattering on lateral sides can affect the impulse components. But on the other hand, the repulsion potential is known [4] to affect charge carriers of a certain sign because of zone bending on the conductor boundaries so that the carriers scatter mirror-like . In work [5] mirror-like scattering of holes in bismuth was revealed by transverse electron focusing. Let the repulsion potential in our case be great and charge carriers scatter mirror-like . Scattering in our two-dimensional structure can also occur due to transverse movements in the direction perpendicular to the planes of the conductor. We shall consider, that nearest quantum level is located highly enough, not lower than 0.1 eV, at least. Energy losses via radiation caused by charge oscillations in the structure, in spite of a high frequency, are hampered. The reason is the following. The spatial amplitude of charge carrier oscillations, related to plasma oscillations, was estimated to be several tenths of angstrom, taking the energy of the oscillator ground state per one carrier equal to $\delta\varepsilon \approx \hbar\omega_s/2N$. This means that the amplitude does not exceed wavelength of charge carriers in the conductor $\lambda_e \approx 1\text{Å}$, characterizing the uncertainty of charge carriers position. In this situation we think, radiation is hampered. In the given structure all oscillations are not possible, satisfying a condition of a resonance. As was stated above, oscillations can have various wave lengths $\lambda_r = 2d/r$. Too small $\lambda$ comparable with the lattice crystallographic constants are hardly possible, because the wave would be sensitive to the boundary roughness and would attenuate fade. For $d = 50$ Å, $\lambda_4 = 25$ Å, which is two times large than a typical lattice constant in the transverse direction to the conducting layer. Therefore, only a few oscillators with small r should be taken into account.

Electron oscillations should not destroy the electron-phonon interaction responsible for Cooper pairing because the amplitude of charge carrier oscillations is small in comparison with $\lambda_e$. Oscillations of a impulse $\delta p$, appropriate $\delta\varepsilon$, are small in comparison with Fermi impulse value $p_F$ and consequently should not worsen interaction between electrons owing to symmetry of a pair wave state on a impulse. Because scattering of electron pairs with opposite directed impulses involving virtual phonons is slow compared to oscillations and is not correlated with them, fast in-time oscillations of pairs impulses due to zero oscillations can be neglected and electronic states can be considered probabilistically smeared in the impulse space. So, zero oscillations affect superconducting properties only through statistics of occupied and free electron pair states.

The number of charge carriers N changes with the conductor size and the dynamic smearing increases proportional to 1/N. It means, that from the point of view of idea of formation of basic BCS state with participation of zero oscillations macroscopic quantum oscillator developed here rigidity of a superconducting state of a conductor with the size d ≈ ξ should be the maximal. Consider now a two-dimensional conductor with a very large d. In the normal state of such a structure, quantum oscillator, similar to the above described cannot exist because the sizes of the conductor are too large and the wave would attenuate. But it is tempting to assume the following. If the Fermi-surface of the conductor is anisotropic, for example, in the form of a square, it would be beneficial for the electronic system on the transition to the superconducting state to disintegrate into a system of domains consistent of space quantum oscillators of the coherence length size, to economy in the kinetic energy by dynamic smearing.

Thus, the work presents arguments for a possible economy in the kinetic energy of electronic pair states in two-dimensional conductors upon their transition to the superconducting state which is connected to the zero oscillations of macroscopic quantum oscillator. The economy can result in a greater rigidity of the superconducting state.

The author thanks V.A. Tulin, V.F. Lukichev, M.Yu. Kupriyanov, G.M. Eliashberg, V.S. Tsoy, V.I. Yudson, V.T. Dolgopolov, I.V. Kukushkin, V.V. Ryazanov, Yu.S. Barash, V.Ya. Kravchenko for discussions. He gratefully acknowledges the support received within framework of the research initiatives "Quantum Macrophysics", "Computing by New Physical Principles" and "Quantum Nanostructures".